\documentstyle[12pt,times]{article}
\input epsf.tex

\newcommand{\nc}{\newcommand}
\newcommand{\rnc}{\renewcommand}

\rnc{\baselinestretch}{1.24}    
\rnc{\arraystretch}{1.24}       

\makeatletter
\rnc{\theequation}{\thesection.\arabic{equation}}
\@addtoreset{equation}{section}
\makeatother

\nc{\be}{\begin{equation}}
\nc{\ee}{\end{equation}}
\nc{\bea}{\begin{eqnarray}}
\nc{\eea}{\end{eqnarray}}
\nc{\xx}{\nonumber\\}

\nc{\eq}[1]{(\ref{#1})}
\nc{\newcaption}[1]{\centerline{\parbox{6in}{\caption{#1}}}}

\nc{\fig}[3]{
\begin{figure}
\centerline{\epsfxsize=#1\epsfbox{#2.eps}}
\newcaption{#3. \label{#2}}
\end{figure}
}

\nc{\np}[3]{Nucl. Phys. {\bf B#1} (#2) #3}
\nc{\pl}[3]{Phys. Lett. {\bf #1B} (#2) #3}
\nc{\prl}[3]{Phys. Rev. Lett.{\bf #1} (#2) #3}
\nc{\prd}[3]{Phys. Rev. {\bf D#1} (#2) #3}
\nc{\ap}[3]{Ann. Phys. {\bf #1} (#2) #3}
\nc{\prep}[3]{Phys. Rep. {\bf #1} (#2) #3}
\nc{\rmp}[3]{Rev. Mod. Phys. {\bf #1} (#2) #3}
\nc{\cmp}[3]{Comm. Math. Phys. {\bf #1} (#2) #3}
\nc{\mpl}[3]{Mod. Phys. Lett. {\bf #1} (#2) #3}
\nc{\cqg}[3]{Class. Quant. Grav. {\bf #1} (#2) #3}
\nc{\jhep}[3]{J. High Energy Phys. {\bf #1} (#2) #3}

\def\CN{{\cal N}}

\def\CN{{\cal N}}

\def\IR{{\hbox{{\rm I}\kern-.2em\hbox{\rm R}}}}
\def\IB{{\hbox{{\rm I}\kern-.2em\hbox{\rm B}}}}
\def\IN{{\hbox{{\rm I}\kern-.2em\hbox{\rm N}}}}
\def\IC{\,\,{\hbox{{\rm I}\kern-.59em\hbox{\bf C}}}}
\def\IZ{{\hbox{{\rm Z}\kern-.4em\hbox{\rm Z}}}}
\def\IP{{\hbox{{\rm I}\kern-.2em\hbox{\rm P}}}}
\def\IH{{\hbox{{\rm I}\kern-.4em\hbox{\rm H}}}}
\def\ID{{\hbox{{\rm I}\kern-.2em\hbox{\rm D}}}}

\def\a{\alpha}
\def\b{\beta}
\def\ga{\gamma}
\def\d{\delta}
\def\ep{\epsilon}

\def\l{\lambda}
\def\m{\mu}
\def\n{\nu}

\def\s{\sigma}

\def\w{\omega}

\def\half{\frac{1}{2}}

\def\del{\nabla}

\def\Tr{{\rm Tr}}

\def\cc{\langle C^{I_1} C^{I_2} \rangle}
\def\vv{\langle V^{I_1} V^{I_2} \rangle}
\def\tt{\langle T^{I_1} T^{I_2} \rangle}

\def\ccc{\langle C^{I_1} C^{I_2} C^{I_3} \rangle}
\def\vcc{\langle V^{I_1} C^{I_2} C^{I_3} \rangle}
\def\tcc{\langle T^{I_1} C^{I_2} C^{I_3} \rangle}

\def\SYMF{{SYM$_4$}}

\nc{\group}[1]{\bigskip\noindent\underline{Group {#1}}\bigskip}

\begin{document}

\begin{titlepage}

\renewcommand{\thefootnote}{\fnsymbol{footnote}}

\begin{flushright}
hep-th/9907108\\
KIAS-P99058\\
\end{flushright}
\vspace*{2.0cm}
\centerline{\Large\bf $\mathbf{AdS_5/CFT_4}$ Four-point Functions of}
\vspace*{0.5cm}
\centerline{\Large\bf Chiral Primary Operators: Cubic Vertices}
\vspace*{1.5cm}
\centerline
{Sangmin Lee \footnote{sangmin@kias.re.kr}}
\vspace*{1.0cm}
\centerline{\sl School of Physics }
\centerline{\sl Korea Institute for Advanced Study}
\centerline{\sl Seoul 130-012 Korea}
\vskip0.3cm
\vspace*{2.0cm}
\centerline{\bf ABSTRACT}
\vspace*{0.5cm}
\noindent
We study the exchange diagrams in the computation of four-point functions of
all chiral primary operators in $D=4$, $\CN=4$ Super-Yang-Mills
using AdS/CFT correspondence.
We identify all supergravity fields that can be exchanged
and compute the cubic couplings.
As a byproduct, we also rederive the normalization of the quadratic action
of the exchanged fields.
The cubic couplings computed in this paper and
the propagators studied extensively in the literature
can be used to compute almost all the exchange diagrams explicitly.
Some issues in computing the complete four-point function in the
``massless sector'' is discussed.

\vspace*{1cm}

\end{titlepage}

\renewcommand{\thefootnote}{\arabic{footnote}}
\setcounter{footnote}{0}

\section{Introduction}

Computing the four-point functions and understanding their structure
is an important problem in the AdS/CFT correspondence
\cite{rmal}-\cite{rev}.
There have been a lot of work on this subject over the past year
\cite{rmuck}-\cite{rskiba}, in which the most extensively studied example
is the $D=4$, $\CN=4$ Super-Yang-Mills theory (\SYMF)
and its string theory dual.
Unlike many other CFTs in the correspondence,
one can compute the correlation functions of
\SYMF\ in the perturbative regime and compare them with the strong coupling
result obtained using the string theory on AdS.
The detailed comparison of two- and three-point functions have shown that
they are not renormalized \cite{rsm, rmit0, rhowe}.
Four-point functions in the two regimes are not expected to be related in a
simple manner, and we hope to learn about non-trivial dynamics of SYM$_4$
in comparing them.

There is a technical difficulty in making such comparison.
On the \SYMF\ side, the easiest four-point functions to compute is those of
the lowest dimensional chiral primary operators $\Tr X^{(i} X^{j)}$
, where $X^i$ are six real scalars of \SYMF \cite{rgps}-\cite{rbkrs}.
On the other hand, in AdS, the easiest supergravity fields turn out to be
the dilaton and the RR scalar, which correspond to
(the supersymmetric completion of)
$\Tr F^2$ and $\Tr F\wedge F$, respectively \cite{rliutsey}-\cite{rsanjay}.
We wish to calculate the four-point functions of $\Tr X^{(i} X^{j)}$
from the AdS side, since it will make a direct comparison between
the two approaches possible.
Note also that since $\Tr F^2$ and $\Tr F\wedge F$
are superconformal descendants of $\Tr X^{(i} X^{j)}$
(one needs to act four supercharges), one can study the relation between
the four-point functions of chiral primaries and those of descendants.
\footnote{
There are convincing arguments which indicate that the
two- or three-point functions of chiral primaries completely determine
those of descendants \cite{rhowe}.
This argument is not valid for four-point functions.
}

The main purpose of this paper is to compute the cubic couplings
that are needed to evaluate the exchange diagrams
for the four-point functions of
{\em arbitrary} chiral primary operators of \SYMF including, of course,
$\Tr X^{(i} X^{j)}$.
Recall that there are two types of connected diagrams contributing to
the four-point functions: exchange and contact.
See Fig. \ref{cubic}.
More specifically, we identify all supergravity fields
that can be exchanged and compute their couplings to
two of the external fields $s^I$ corresponding to
the chiral primary operators.
As a byproduct, we also determine the normalization of their
quadratic action (thereby normalizing their propagator properly).

To evaluate the exchange diagrams, we also need the propagators and
have to integrate over the location of the vertices in AdS.
Fortunately, almost everything about the propagators and the integrals are
known in the literature \cite{rmit2}-\cite{rmit6},
with the propagator of a massive symmetric tensor particle
being the only missing piece for our purpose.

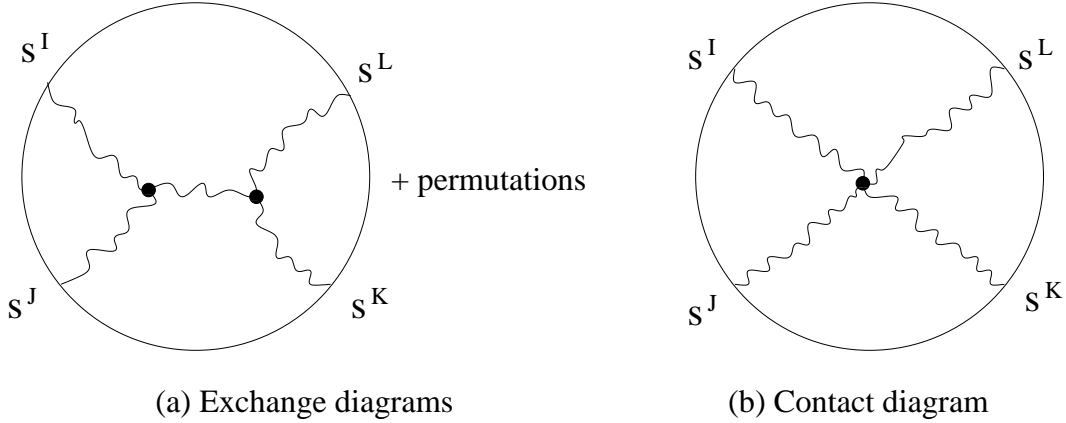
\fig{400pt}{cubic}{The two types of supergravity diagram contributing to
the AdS computation of four-point functions of chiral primary operators.}

At the end of this paper, we get back to our primary concern and restrict
our attention to $\Tr X^{(i} X^{j)}$ and their superconformal decendants.
Most physically interesting operators belong to this group, which
form the massless multiplet of the $SU(2,2|4)$ superconformal algebra.
\footnote{The term ``massless'' comes from the fact that their supergravity
dual all belong to the same supermultiplet as the massless graviton.
Since mass is not a Casimir of the AdS supersymmetry, most of the fields
in this multiplet has non-zero AdS mass.}
It is known that the supergravity on $AdS_5$ can be truncated to contain
the massless multiplet only \cite{rgauge1, rgauge2, rtrunc}.
We discuss how this fact can be exploited in computing the complete
four-point function of the (not necessarily chiral) operators in the
massless multiplet.

\bigskip

As this work was being completed, we received reference \cite{rafv}
which computes the same cubic couplings.


\section{Identification of the Exchanged Fields}

We wish to pick out the Kaluza-Klein modes of Type IIB supergravity on
$AdS_5\times S^5$ \cite{rkim, rgunaydin} which
have non-zero cubic coupling with two $s^I$.
Obviously, there cannot be any cubic coupling of
a single fermion with two bosons,
so we only need to examine the bosonic modes.

We note that $s^I$ are singlets under the $SL(2,\IR)$ symmetry group
of the Type IIB supergravity. This means that two $s^I$ can have a cubic
coupling with a third field only the latter is again a singlet.
In fact, it is sufficient to consider the $U(1)_Y$ subgroup of $SL(2,\IR)$.
\footnote{
It was argued in \cite{rintr2} that the $U(1)_Y$ symmetry goes beyond the
supergravity approximation and becomes an exact symmetry of the operator
product expansions for which at least two of the three operators are short.
This is the case for our discussion.
}
In $D=10$, the RR-scalar, the dilaton and the two two-form fields have
non-zero charge under the $U(1)_Y$ and the same is true of all their
Kaluza-Klein modes. Hence they do not couple to $s^I$ at cubic order.

The graviton and the RR four-form potential in $D=10$ are $U(1)_Y$ singlets,
and their Kaluza-Klein modes may couple to $s^I$. It turns out that
all but two Kaluza-Klein towers have non-zero cubic couplings.
Before explaining the reason for the exceptions,
we need to recall the defintion of the modes
following references \cite{rkim}.
After fixing a gauge and solving for the constraints, the spherical
harmonics decomposition of the physical modes read
\be
\begin{array}{rclrcl}
h^\a_\a &=& h_2^I Y^I, &
a_{\a_1 \a_2 \a_3 \a_4} &=& b^I \ep_{\a_1 \a_2 \a_3 \a_4 \a} \del^\a Y^I \\
h_{\m\a} &=& h_\m^I Y_\a^I &
a_{\m\a_1\a_2\a_3} &=& \phi_\m^I\ep_{\a_1\a_2\a_3}{^{\b\ga}}\del_\b Y^I_\ga \\
h_{(\a\b)} &=& \phi^I Y^I_{(a\b)}, &
a_{\m\n\a\b} &=& b_{\m\n}^{I,\pm} Y_{[\a\b]}^{I,\pm} \\
h'_{(\m\n)} &=& H_{(\m\n)}^I Y^I, & & & \\
\end{array}
\ee
Following \cite{rsm}, we define two mass eigenstate scalars
\be
s^I = {1\over 20(k+2)} \{ h_2^I - 10(k+4)b^I \}, \  \
t^I = {1\over 20(k+2)} \{ h_2^I + 10k b^I \}.
\ee
As in \cite{rgunaydin}, we also define two mass eigenstate vectors
\be
V_\m = h_\m - 4(k+3)\phi_\m, \  \
W_\m = h_\m +4(k+1) \phi_\m.
\ee
The scalar $\phi^I$, symmetric traceless tensor $H_{(\m\n)}^I$ and
the two-forms $b_{\m\n}^{I,\pm}$ are mass eigenstates on their own.

Among these Kaluza-Klein towers, $b_{\m\n}^{I,\pm}$ are the only two
that decouple from $s^I$ completely. The reason is that the Clebsch-Gordan
coefficient of two $s^I$ and a $b_{m\n}^{I,\pm}$ for the
$SU(4)=\mbox{Spin}(6)$ $R$-symmetry group is zero at all levels.
The $R$-symmetry quantum number of the
antisymmetric tensor spherical harmonics
$Y_{[\a\b]}^{I,+}$ and $Y_{[\a\b]}^{I,-}$ are
$(k+1,k-1,0)$ and $(k+1,k+1,2)$, respectively,
where the $i$-th number in each triplet is the number of
boxes on the $i$-th row of an $SU(4)$ Young tableaux.
On the other hand, the $R$ symmetry quantum number of $s^I$ is $(k,k,0)$.
Using the standard tensor-product rule in terms of Young tableaux,
one can show that the tensor product of $(k_1,k_1,0)$ and $(k_2,k_2,0)$
do not contain $(k_3+1,k_3-1,0)$ or $(k_3+1,k_3+1,2)$ for any $k_3$.


\section{Quadratic Action and Cubic Couplings}

\subsection{Quadratic Action}

The part of the $D=5$ action that is relevant to our discussion can be
written as
\be
\label{action}
S = {4N^2 \w_5 \over (2\pi)^5} \int d^5 x \sqrt{-g}
\left\{ L_2 + L_3 \right\},
\ee
where the constant in front the integral is (the inverse of) the
five dimensional Newton's constant including the volume of a unit
5-sphere $\w_5 = \pi^3$.
The integrands $L_2$ and $L_3$ are the quadratic and cubic terms of the
Lagrangian. The quadratic Lagrangian takes the following form.
\be
\label{actionquad}
\begin{array}{rcl}
2 L_2 &=&
- A^1_I \{ (\del s^I)^2 + (m^1_I)^2 (s^I)^2 \}
- A^2_I \{ (\del t^I)^2 + (m^2_I)^2 (t^I)^2 \}
- A^3_I \{ (\del \phi^I)^2 + (m^3_I)^2 (\phi^I)^2 \}
\\
&&- A^4_I \left\{ \half F^2_{\m\n}(V^I)
+(m^4_I)^2 (V^I_\m)^2 \right\}
- A^5_I \left\{ \half F^2_{\m\n}(W^I)
+ (m^5_I)^2 (W^I_\m)^2 \right\}
\\
&&- A^6_I \left\{ (\del_\l H^I_{(\m\n)})^2 + (m^6_I)^2 (H^I_{(\m\n)})^2
\right\},
\end{array}
\ee
where $F_{\m\n}(V^I) \equiv \del_\m V^I_\n - \del_\n V^I_\m$ and
similarly for $F_{\m\n}(W^I)$. The normalization constants and the
masses are given by
\be\label{quadnorm}
\begin{array}{ll}
A^1_I = \displaystyle
32 {k(k-1)(k+2) \over k+1} z(k),    &\  \ (m^1_I)^2 = k(k-4),\xx
A^2_I = \displaystyle
32 {(k+4)(k+5)(k+2) \over k+3} z(k),&\  \ (m^2_I)^2 = (k+4)(k+8),\xx
A^3_I = \displaystyle
\half z(k),    &\  \ (m^3_I)^2 = k(k+4),\xx
A^4_I = \displaystyle
{k+1 \over 2(k+2)} z(k),&\  \ (m^4_I)^2 = (k-1)(k+1),\xx
A^5_I = \displaystyle
{k+3 \over 2(k+2)} z(k),&\  \ (m^5_I)^2 = (k+3)(k+5),\xx
A^6_I = \displaystyle
\half z(k),&\  \ (m^6_I)^2 = k(k+4) - 2.
\end{array}
\ee
The normalization of the quadratic action for $s^I$ was first computed in
\cite{rsm}. Subsequent papers \cite{raru, richard} pointed out some flaws
in the original derivation, but confirmed that the answer was correct.
The normalzation constants for other fields were first computed in
\cite{raru}.
As will be explained in subsection \ref{revisit},
our computation of the cubic couplings provide another way to fix the
normalizations.

\subsection{Cubic Couplings: Results}

As we will show in the next section,
the cubic Lagrangian in \eq{action} is given by
\bea\label{actioncubic}
2 L_3 &=&
- {1 \over 3} G^1_{IJK} s^I s^J s^K - G^2_{IJK} t^I s^J s^K
- G^3_{IJK} \phi^I s^J s^K
\xx
&&- G^4_{IJK} V^I_\m s^J \del^\m s^K
- G^5_{IJK} W^I_\m s^J \del^\m s^K
- G^6_{IJK} H^I_{(\m\n)} s^J \del^{(\m}\del^{\n)}s^K,
\eea
where the coupling constants are defined by
\bea\label{cubiccoup}
G^1_{I_1I_2I_3} &=&
2^9 {\s(\s^2-1)(\s^2-4)\a_1\a_2\a_3 \over (k_1+1)(k_2+1)(k_3+1)}
a(k_1,k_2,k_3) \ccc,
\xx
G^2_{I_1I_2I_3} &=& - 2^8{(\s+2) \a_1! (\a_2+2)(\a_3+2)
\over (\a_1-5)! (k_1+3)(k_2+1)(k_3+1)}
a(k_1,k_2,k_3) \ccc,
\xx
G^3_{I_1I_2I_3} &=&
-2^5 {\s(\s+1)(\a_1-1)(\a_1-2) \over (k_2+1)(k_3+1)}
h(k_1,k_2,k_3) \tcc,
\xx
G^4_{I_1I_2I_3} &=&
2^5 {(k_1+1)(\s^2-{1\over 4})(\s+{3\over 2})(\a_1-\half)
\over (k_1+2)(k_2+1)(k_3+1)}
e(k_1,k_2,k_3) \vcc,
\xx
G^5_{I_1I_2I_3} &=&
2^5 {(k_1+3)(\s+{3\over 2})(\a_1-\half)(\a_1-{3\over2})(\a_1-{5\over2})
\over (k_1+2)(k_2+1)(k_3+1)}
e(k_1,k_2,k_3) \vcc,
\xx
G^6_{I_1I_2I_3} &=&
2^5 {(\s+1)(\s+2)\a_1(\a_1-1)\over (k_2+1)(k_3+1)}
a(k_1,k_2,k_3) \ccc.
\eea
The definitions of the symbols $\s, \a_{1,2,3}$, the functions
$a(k_1,k_2,k_3)$, $e(k_1,k_2,k_3)$, $h(k_1,k_2,k_3)$
and the brackets $\ccc$, $\vcc$, $\tcc$
are given in Appendix A.

\subsection{Cubic Couplings: Derivation}

We use the method developed in \cite{rsm} to calculate the cubic couplings.
The starting point is the quadratic corrections to the field equation for
$s^I$,
\be\label{eomqs}
\{ \del^2 - (m^1_I)^2 \} s^I = {1\over 2(k+2)}
\left\{
(k+4)(k+5)Q_1 + Q_2 + (k+4) (\del_\m Q_3^\m + Q_4)\right\}^I,
\ee
where $Q_{1,2,3,4}$ are defined in equations (3.24) and (3.25) of \cite{rsm}.
To calculate the coupling of two $s^I$ with another field, say $\Psi$,
one looks for terms proportional to $s\Psi$ on the right-hand-side of
\eq{eomqs}. The collection of such terms usually contain higher derivative
terms of $s^I$ which can be removed by a field redefinition.
The final product of the calculation takes the form
\be\label{almostcubic}
\begin{array}{rcl}
\{ \del^2 - (m^1_I)^2 \} s^I &=&
\half \l^1_{IJK} s^J s^K + \l^2_{JIK} t^J s^K + \l^3_{JIK} \phi^J s^K \\
& & + \l^4_{JIK} V^J_\m \del^\m s^K + \l^5_{JIK} W^J_\m \del^\m s^K
+ \l^6_{JIK} H^J_{(\m\n)} \del^{(\m}\del^{\n)}s^K.
\end{array}
\ee
If we multiply it by the normalization constant $A^1_I$, we obtain
\be\label{eomcubic}
\begin{array}{rcl}
A^1_I \{ \del^2 - (m^1_I)^2 \} s^I &=&
\half G^1_{IJK} s^J s^K + G^2_{JIK} t^J s^K + G^3_{JIK} \phi^J s^K \\
& & + G^4_{JIK} V^J_\m \del^\m s^K + G^5_{JIK} W^J_\m \del^\m s^K
+ G^6_{JIK} H^J_{(\m\n)} \del^{(\m}\del^{\n)}s^K.
\end{array}
\ee
Remarkably, all the coupling constants $G^n_{IJK}$ satisfy non-trivial
conditions which ensures that the equation of motion can be derived from
an action of the form \eq{action}, \eq{actionquad}, \eq{actioncubic}.
In the following, we briefly sketch the intermediate steps of the
calculation that lead to \eq{almostcubic} for each coupling.

\subsubsection{Coupling to Scalars: $sss$, $sst$, $ss\phi$}

We recall the derivation of $sss$ vertices from \cite{rsm}.
Collecting all the terms that are quadratic
in $s$ on the right-hand-side of \eq{eomqs}, one finds
\be
\label{alsss}
\{ \del^2 - (m^1_I)^2 \} s^I =
D_{IJK}s^Js^K + E_{IJK} \del_\m s^J \del^\m s^K
+F_{IJK} \del^{(\m}\del^{\n)}s^J \del_{(\m}\del_{\n)}s^K,
\ee
where $D$, $E$ and $F$ are some functions of $I$, $J$ and $K$.
One can remove the derivative terms on the right-hand-side of \eq{alsss}
by a field redefinition
\be
\label{redeff1}
s^I = {s'}^I + J_{IJK} {s'}^J {s'}^K + L_{IJK} \del^\m {s'}^J \del_\m {s'}^K,
\ee
where
\be
\label{redeff2}
L_{IJK} = \half F_{IJK}, \  \
J_{IJK} = \half E_{IJK}
+ {1\over 4} F_{IJK} \{ (m^1_I)^2 -(m^1_J)^2 - (m^1_K)^2 + 8)
\ee
such that \eq{alsss} takes the form of \eq{almostcubic} with
\be
\l^1_{IJK} = D_{IJK} - \{ (m^1_J)^2 + (m^1_K)^2 -(m^1_I)^2 ) J_{IJK}
-{2\over 5} L_{IJK} (m^1_J)^2 (m^1_K)^2
\ee

In order to compute the $sst$ and $ss\phi$ vertices, one collect the terms
proportional to $st$ or $s\phi$ from \eq{eomqs}.
The $st$ and $s\phi$ terms appear on the right-hand-side of \eq{alsss}
with the same number of derivatives.
The derivative terms can be removed by a field redefinition similar to
\eq{redeff1}, \eq{redeff2} with
appropriate values of the masses for different fields.

\subsubsection{Coupling to Vectors: $ssV_\m$, $ssW_\m$}

The terms proportional to $sV_\m$ and $sW_\m$ on the right-hand-side of
\eq{eomqs} add up to yield
\be
\begin{array}{rcl}
\{ \del^2 - (m^1_I)^2 \} s^I &=&
D^4_{JIK}V^J_\m \del^\m s^k + D^5_{JIK}W^J_\m \del^\m s^k \\
& & +E^4_{IJK} \del_\m V^J_\n \del^\m \del^\n s^K
+E^5_{IJK} \del_\m W^J_\n \del^\m \del^\n s^K,
\end{array}
\ee
Consider the field redefinition
\bea
s^I = {s'}^I
+ J^4_{JIK} V^J_\m \del^\m {s'}^K + J^5_{JIK} W^J_\m \del^\m {s'}^K
\eea
To remove the $E_{JIK}$ terms, we set $J_{JIK} = \half E_{JIK}$.
Then, in terms of the redefined field, the equation of motion reads
\be
\{ \del^2 - (m^1_I)^2 \} s^I =
\l^4_{JIK}V^J_\m \del^\m s^k + \l^5_{JIK}W^J_\m \del^\m s^k,
\ee
where
\be
\l^{4,5}_{JIK} = D^{4,5}_{JIK} -
\half \{ (m^{4,5}_J)^2 + (m^1_K)^2 - (m^1_I)^2 - 8 \}.
\ee

\subsubsection{Coupling to Symmetric Tensors: $ssH_{(\m\n)}$}

The same method is applicable here, except that there is no need to
make a field redefinition.

\subsection{Quadratic Action Revisited \label{revisit}}

In the previous subsection, we obtained the coupling constants
using the equation of motion of $s^I$ only.
One could use the field equations for other fields to
derive the same coupling.
For example, the quadratic correction to the equation for $t^I$
will contain a term like
\be
\{ \del^2 - (m^2_I)^2 \} t^I = \half\tilde{\l}_{IJK} s^J s^K
\ee
after the same kind of field redefinition we made before.
In order for an action of the type \eq{action} to exist,
the normalization of the $t^I$ kinetic term must satisfy
\be
A^2_I \tilde{\l}_{IJK} = G^2_{IJK}.
\ee
where the $G^2_{IJK}$ was obtained in \eq{eomcubic}.
This means that one can determine $A^2_I$ by comparing
the two equations of motion if $A^1_I$ is known.

One can use this method to determine the normalization constants for
all fields from that of $s^I$.
One can also reverse the logic and determine the normalization constant of
$s^I$ from that of some other field. For example, the Kaluza-Klein tower
$H^I_{(\m\n)}$ contains the $D=5$ graviton and its massive counterparts.
On a general ground, even though a simple form of the covariant action in
$D=10$ is not available, one expect that the action in $D=5$ will take
precisely the form
\be
S = {V \over 2\kappa_{10}^2} \int d^5x \sqrt{-g} \{ R + \cdots \},
\ee
where $V$ is the volume of the internal space and $R$ is the curvature of
the 5-dimensional metric. This fixes the normalization for $H^I_{(\m\n)}$
completely, and one can use it to determine those for other fields.
We have rederived all the normalization constants this way, compared them
with the previous results in \cite{raru} and found perfect agreement.
The details of the computation is not particularly illuminating and
we omit it here.


\section{On the Massless Multiplet}

In view of the fact that most of work on the four-point functions
deal with the operators in the massless multiplet, it is of particular
interest to examine the couplings of $s^{(ij)}$ corresponding to the operator
$\Tr X^{(i} X^{j)}$. Consider the cubic couplings in \eq{cubiccoup} for
$k_2=k_3=2$. From the way the $SO(6)$ tensors are contracted, one finds that
the couplings can be non-zero only for $k_1 \le 4$.
It is easy to compute the couplings for all such $k_1$ to find that
the only non-vanishing couplings are $G^4_{I_1I_2I_3}$ for $k_1=1$
and $G^6_{I_1I_2I_3}$ for $k_1=0$.
{}From the mass formula in \eq{quadnorm}, one notices that the
supergravity fields that couple to two $s^{(ij)}$ are a massless vector and
the graviton. They correspond to the R-symmetry current and
the energy-momentum tensor operators of \SYMF, respectively.

This should not come as a surprise. It is known that the Type IIB supergravity
on $AdS_5\times S^5$ can be consitently truncated to contain the massless
multiplet. The resulting theory is identified with the $D=5$, $\CN=8$,
$SO(6)$ gauged supergravity \cite{rgauge1, rgauge2, rtrunc}.
The fields $t^I$, $\phi^I$ and $W^I_\m$ do not have a component in the
massless sector, so in order for the truncation to be possible,
they should not couple to two $s^{ij}$.
Our computation explicitly shows that it is indeed the case.

One can make a step further. In the gauged supergravity, the coupling
of $s^{(ij)}$ with the vector and the graviton at cubic order is completely
determined by gauge invariance and general covariance. That is, the action
must take the form (we suppress the paranthesis in $s^{(ij)}$ below)
\be
\label{actiongauge}
S = {1\over 2\kappa_5^2} \int d^5x \sqrt{-g}
\left[ R + 12 - {1\over 8g^2} (F^{ij}_{\m\n})^2
- {1\over 4} \{ (D s^{ij})^2 - 4 (s^{ij})^2 \} + \cdots \right],
\ee
where the gauge-covariant derivative is defined by
\be
\label{covder}
D_\m s^{ij} = \del_\m + V_\m^{ik} s^{kj} + V_\m^{jk} s^{ki},
\ee
and $V^{ij}_\m$ are real and antisymmetric tensors of $SO(6)$.
$F^{ij}_{\m\n}$ are the non-abelian field strength.

We want to compare \eq{actiongauge} with the massless sector of
\eq{action}, \eq{actionquad}, \eq{quadnorm}, \eq{actioncubic}
and \eq{cubiccoup}. First, we can absorb $A^1_I$ into the normalization of
$s^{ij}$ to bring the kinetic term into the form \eq{actiongauge}.
General covariance of \eq{actiongauge} then uniquely fixes the $ssH_{(\m\n)}$
coupling, since it comes from the expansion of the metric multiplying
the kinetic term. Not surprisingly, the result agree with \eq{cubiccoup}.
Next, we change the normalization of $V_\m$ such that the cubic coupling
in \eq{cubiccoup} is identified with that derived by expanding the covariant
derivative \eq{covder}. This fixes the value of Yang-Mills coupling constant
$g^2$ in \eq{actiongauge}. Keeping track of numerical factors carefully,
one finds that $g^2 = 4$. This is in agreement with an independent analysis
of \cite{rtrunc}.

Note that, contrary to the complicated derivation of the cubic couplings from
the Kaluza-Klein reduction of Type IIB supergravity,
the action of the gauged supergravity \eq{actiongauge} is rather simple.
Moreover, the complete action and supersymmetry transformation rules are known
\cite{rgauge1, rgauge2}. Therefore, in order to compute the correlation
functions of operators in the massless multiplet only, it may be easier to
start from the gauged supergravity rather than the full $D=10$ supergravity.
This is perhaps the right way to tackle the formidable task
of computing the quartic vertices if one is interested
in the four-point functions of the operators in the massless multiplet only.

\vskip 1cm
\centerline{\bf ACKNOWLEDGEMENT}
\vskip 0.5cm

\noindent
I am grateful to Mukund Rangamani for collaboration at an early stage of this
work and Shiraz Minwalla for discussions.


\newpage
\centerline{\Large\bf Appendix}
\vskip 1cm
\appendix
\section{Spherical Harmonics Integrals}

\subsection{Scalar, Vector and Tensor Spherical Harmonics}

A self-contained introduction to spherical harmonics is given in \cite{rsm}.
Here we summarize the defining properties of spherical harmonics on $S^5$.

\begin{itemize}

\item
Scalar spherical harmonics $Y^I$ are defined to be the functions in $\IR^6$
of the form
\be
Y^I = C^I_{i_1 \cdots i_k} x^{i_1} \cdots x^{i_k},
\ee
restricted to a unit $S^5$ centered at the origin.
$x^i$ are the Cartesian coordintes of $\IR^6$ and $C^I$
are totally symmetric and traceless tensors of $SO(6)$.

\item
Vector spherical harmonics $Y^I_\a$ are defined to be the tangent components
of the vector field in $\IR^6$ of the form
\be
Y^I_a = (V^I)^a_{i_1 \cdots i_k} x^{i_1} \cdots x^{i_k},
\ee
where the tensor $(V^I)^a_{i_1 \cdots i_k}$ are symmetric and traceless
in $i_1,\cdots, i_k$ and vanishes when symmetrized over $a$ and all $i$s.

\item
Tensor spherical harmonics $Y^I_{(\a\b)}$ are defined to be the
tangents component of the tensor field in $\IR^6$ of the form
\be
Y^I_{ab} = (T^I)^{ab}_{i_1 \cdots i_k} x^{i_1} \cdots x^{i_k},
\ee
where the tensor $(T^I)^{ab}_{i_1 \cdots i_k}$ are symmetric and traceless
in $i_1,\cdots, i_k$ and $a,b$ and vanishes when symmetized over
$b$ and all $i$s with fixed $a$.

\end{itemize}

\subsection{Quadratic Integrals}

The following integrals of the spherical harmonics are needed in the
calculation of the main text.
\be
\begin{array}{rcl}
\label{sphintquad}
\
{1\over \w_5}
\int Y^{I_1} Y^{I_2} &=& z(k) \cc,
\\
{1\over \w_5}
\int \del_\a Y^{I_1} \del^\a Y^{I_2} &=&
f(k)z(k) \cc,
\\
{1\over \w_5}
\int \del_{(\a}\del_{\b)}Y^{I_1} \del^{(\a}\del^{\b)}Y^{I_2} &=&
q(k)z(k) \cc,
\\
{1\over \w_5}
\int Y^{I_1}_\a Y^{I_2}_\b g^{\a\b} &=& z(k) \vv,
\\
{1\over \w_5}
\int Y^{I_1}_{(\a\b)} Y^{I_2}_{(\ga\d)} g^{\a\ga} g^{\b\d} &=& z(k) \tt,
\end{array}
\ee
where we normalized the integral by the volume of a unit 5-sphere
$\w_5 = \pi^3$. The brackets in the formulas are defined by
\be
\begin{array}{rcl}
\cc &\equiv& C^{I_1}_{i_1 \cdots i_k} C^{I_2}_{i_1 \cdots i_k}, \\
\vv &\equiv& (V^{I_1})^a_{i_1 \cdots i_k} (V^{I_2})^a_{i_1 \cdots i_k},\\
\tt &\equiv& (T^{I_1})^{ab}_{i_1 \cdots i_k} (T^{I_2})^{ab}_{i_1 \cdots i_k},
\end{array}
\ee
The functions $z$, $f$ and $q$ are defined by
\be
\label{quadint}
\begin{array}{rcl}
z(k) &=& {1 \over (k+1)(k+2) 2^{k-1}}, \\
f(k) &=& k(k+4),            \\
q(k) &=& {4\over 5}k(k-1)(k+4)(k+5),
\end{array}
\ee
The method of deriving \eq{sphintquad} is explained in \cite{rsm}.


\subsection{Cubic Integrals}

We also need the following cubic integrals of spherical harmonics,
\be
\label{cubintdef}
\begin{array}{rcl}
{1\over \w_5} \int Y^{I_1} Y^{I_2} Y^{I_3}
&=& a(k_1,k_2,k_3) \ccc,
\\
{1\over \w_5} \int Y^{I_1} \del_{\a}Y^{I_2} \del^{\a}Y^{I_3}
&=& b(k_1,k_2,k_3) \ccc,
\\
{1\over \w_5} \int \del^{(\a}\del^{\b)}Y^{I_1}\del_{\a}Y^{I_2}\del_{\b}Y^{I_3}
&=& c(k_1,k_2,k_3) \ccc,
\\
{1\over \w_5}\int Y^{I_1} \del^{(\a}\del^{\b)}Y^{I_2}\del_{\a}\del_{\b}Y^{I_3}
&=& d(k_1,k_2,k_3) \ccc,
\\
{1\over \w_5}\int Y_\a^{I_1}Y^{I_2}\del^\a Y^{I_3}
&=& e(k_1,k_2,k_3) \vcc,
\\
{1\over \w_5}\int Y_\a^{I_1} \del_\b Y^{I_2} \del^{(\a}\del^{\b)} Y^{I_3}
&=& f(k_1,k_2,k_3) \vcc,
\\
{1\over \w_5}\int \del_\a Y_\b^{I_1} Y^{I_2} \del^{(\a}\del^{\b)} Y^{I_3}
&=& g(k_1,k_2,k_3) \vcc,
\\
{1\over \w_5}\int Y_{(\a\b)}^{I_1} \del^\a Y^{I_2} \del^\b Y^{I_3}
&=& h(k_1,k_2,k_3) \tcc,
\\
{1\over \w_5}
\int \del_\ga Y_{(\a\b)}^{I_1} \del^{(\a}\del^{\b)} Y^{I_2} \del^\ga Y^{I_3}
&=& i(k_1,k_2,k_3) \tcc.
\end{array}
\ee
The brackets are defined by
\be
\begin{array}{rcl}
\ccc &\equiv&
C^{I_1}_{\{\a_2\}\{\a_3\}}
C^{I_2}_{\{\a_3\}\{\a_1\}}C^{I_3}_{\{\a_1\}\{\a_2\}}, \\
\vcc &\equiv&
{1 \over \a_3 +\half} (V^{I_1})^a_{\{\a_2-1/2\}\{\a_3+1/2\}}
C^{I_2}_{\{\a_3+1/2\}\{\a_1-1/2\}}C^{I_3}_{a\{\a_2-1/2\}\{\a_1-1/2\}}
\\ &=&
- \langle V^{I_1} C^{I_3} C^{I_2} \rangle, \\
\tcc &\equiv&
(T^{I_1})^{ab}_{\{\a_2\}\{\a_3\}}
C^{I_2}_{a\{\a_3\}\{\a_1-1\}}C^{I_3}_{b\{\a_2\}\{\a_1-1\}},
\end{array}
\ee
where we defined
\be
\s = \half(k_1+k_2+k_3),\  \  \a_i = \s - k_i \  \ (i=1,2,3).
\ee
The functions are defined by
\be
\label{cubintans}
\begin{array}{rcl}
a(k_1,k_2,k_3) &=& {1\over (\s +2)!2^{\s-1}}
{k_1!k_2!k_3! \over \a_1!\a_2!\a_3!}, \\
b(k_1,k_2,k_3) &=& \half ( f_2 + f_3 - f_1 ) a(k_1,k_2,k_3),    \\
c(k_1,k_2,k_3) &=&
{1 \over 4} ( f_2^2 -f_1^2 -f_3^2 +2 f_2 f_3) a(k_1,k_2,k_3),
+ {1\over 5} f_1 b(k_1,k_2,k_3) \\
d(k_1,k_2,k_3) &=& \half ( f_2 + f_3 - f_1 - 8 ) b(k_1,k_2,k_3)
- {1 \over 5} f_2 f_3 a(k_1,k_2,k_3).
\\
e(k_1,k_2,k_3) &=&
{1\over (\s + {3\over 2})!2^{\s-{3\over 2}}}
{k_1!k_2!k_3! \over (\a_1-\half)!(\a_2-\half)!(\a_3-\half)!}, \\
f(k_1,k_2,k_3) &=&
\left\{ \half (f_2+f_3-f_1-3) -{1\over 5} f_3 \right\} e(k_1,k_2,k_3), \\
g(k_1,k_2,k_3) &=&
\left\{ \half (f_1+f_3-f_2-5) -{1\over 5} f_3 \right\} e(k_1,k_2,k_3).
\\
h(k_1,k_2,k_3) &=& 2(\s +2)\a_1 a(k_1,k_2,k_3), \\
i(k_1,k_2,k_3) &=& \half (f_2-f_1-f_3-8) h(k_1,k_2,k_3).
\end{array}
\ee
Again, these formulas can be derived using the method of \cite{rsm}.



\newpage
\begin{thebibliography}{99}


\bibitem{rmal}
J. Maldacena,
``The Large $N$ limit of Superconformal Field Theories and Supergravity,''
hep-th/9711200.

\bibitem{rwit}
E. Witten, ``Anti-de Sitter Space and Holography,''
hep-th/9802150.

\bibitem{rpol}
S. S. Gubser, I. R. Klebanov and A. M. Polyakov,
``Gauge Theory Correlators from Non-critical String Theory,''
\pl{428}{1998}{105}, hep-th/9802109.

\bibitem{rev}
For a review, see \\
O. Aharony, S. S. Gubser, J. Maldacena, H. Ooguri and Y. Oz,
``Large $N$ Field Theories, String Theory and Gravity,''
hep-th/9905111.




\bibitem{rmuck}
W. M\"uck and  K. S. Viswanathan,
``Conformal Field Theory Correlators from Classical Field Theory
on Anti-de Sitter Space II. Vector and Spinor Fields,''
hep-th/9805145.

\bibitem{rliutsey}
H. Liu and A. A. Tseytlin,
``On Four-point Functions in the CFT/$AdS$ Correspondence,''
hep-th/9807097.

\bibitem{rmit1}
D. Z. Freedman, S. D. Mathur, A. Matusis and L. Rastelli,
``Comments on 4-point Functions in the CFT/$AdS$ Correspondence,''
hep-th/9808006.

\bibitem{rjohnfour}
J. H. Brodie and M. Gutperle,
``String Corrections to Four-point Functions in the $AdS$/CFT Correspondence,''
hep-th/9809067.

\bibitem{rmit2}
E. D'Hoker and D. Z. Freedman,
``Gauge Boson Exchange in $AdS_{d+1}$,''
hep-th/9809179.

\bibitem{rhliu}
H. Liu,
``Scattring in Anti-de Sitter Space and Operator Product Expansion,''
hep-th/9811152.

\bibitem{rmit3}
E. D'Hoker and D. Z. Freedman,
``General Scalar Exchange in $AdS_{d+1}$,''
hep-th/9811257.

\bibitem{rmit4}
E. D'Hoker, D. Z. Freedman, S. D. Mathur, A. Matusis and L. Rastelli,
``Graviton and Gauge Boson Propagators in $AdS_{d+1}$,''
hep-th/9902042.

\bibitem{rmit5}
E. D'Hoker, D. Z. Freedman, S. D. Mathur, A. Matusis and L. Rastelli,
``Graviton Exchange and Complete 4-point Functions in the AdS/CFT
correspondence,''
hep-th/9903196.

\bibitem{rmit6}
E. D'Hoker, D. Z. Freedman and L. Rastelli,
``ADS/CFT 4-point Functions: How to Succeed at $z$-integrals without Really
Trying,''
hep-th/9905049.

\bibitem{rsanjay}
Sanjay,
``On Direct and Crossed Channel Asymptotics of Four-Point Functions in
AdS/CFT correspondence,''
hep-th/9906099.




\bibitem{rgps}
F. Gonzalez-Rey, I. Park, K. Schalm,
``A Note on Four-Point Functions of Conformal Operators in $\CN=4$
Super-Yang-Mills,''
hep-th/9811155.

\bibitem{rehssw1}
B. Eden, P. S. Howe, C. Schubert, E. Sokatchev and P. C. West,
``Four-point Functions in $\CN=4$ Supersymmetric Yang-Mills Theory at
Two Loops,''
hep-th/9811172.

\bibitem{rehssw2}
B. Eden, P. S. Howe, C. Schubert, E. Sokatchev and P. C. West,
``Simplifications of Four-Point Functions in $\CN=4$ Supersymmetric
Yang-Mills theory at Two Loops,''
hep-th/9906051.

\bibitem{rbkrs}
M. Bianchi, S. Kovacs, G. Rossi and Y. S. Stanev,
``On the Lograithmic Behaviour in $\CN=4$ SYM theory,''
hep-th/9906188.

\bibitem{rskiba}
W. Skiba,
``Correlators of Short Multi-Trace Operators in $N=4$
Supersymmetric Yang-Mills,''
hep-th/9907088.


\bibitem{rsm}
S. Lee, S. Minwalla, M. Rangamani and N. Seiberg,
``Three-Point Functions of Chiral Operators in $D=4$, $\CN=4$ SYM
at Large $N$,''
hep-th/9806074.

\bibitem{rmit0}
E. D'Hoker, D. Z. Freedman and W. Skiba,
``Field Theory Tests for Correlators in the $AdS$/CFT Correspondence,''
hep-th/9807098.

\bibitem{rhowe}
P.S. Howe, E. Sokatchev and P.C. West,
``3-Point Functions in $\CN=4$ Yang-Mills,''
\pl{444}{1998}{341}, hep-th/9808162.



\bibitem{rintr}
K. Intriligator,
``Bonus Symmetries of $\CN=4$ Super-Yang-Mills Correlation Functions
via AdS Duality,''
\np{551}{1999}{575}, hep-th/9811047.

\bibitem{rintr2}
K. Intriligator and W. Skiba,
``Bonus Symmetry and the Operator Product Expansion of $\CN=4$
Super-Yang-Mills,''
hjep-th/9905020.



\bibitem{rkim}
H. J. Kim, L. J. Romans and P. van Nieuwenhuizen
``Mass Spectrum of Chiral Ten-dimensional $\CN=2$ Supergravity on $S^5$,''
\prd{32}{1985}{389}.

\bibitem{rgunaydin}
M. G\"unaydin and N. Marcus,
``The Spectrum of the $S^5$ Compactification
of the Chiral $\CN=2$, $D=10$ Supergravity
and the Unitary Supermultiplets of $U(2,2|4)$,''
\cqg{2}{1985}{L11}.



\bibitem{raru}
G. E. Arutyunov and S. A. Frolov,
``Quadratic Action for Typ IIB Supergravity on $AdS_5 \times S^5$,''
hep-th/9811106.

\bibitem{richard}
R. Corrado, B. Florea and R. McNees,
``Correlation Functions of Operators and Wilson Surfaces in the $d=6$,
$(0,2)$ Theory in the Large $N$ Limit,''
hep-th/9902153.



\bibitem{rgauge1}
M. G\"unaydin, L. J. Romans and N. P. Warner,
``Gauged $\CN=8$ Supergravity in Five Dimensions,''
\pl{154}{1985}{268}.

\bibitem{rgauge2}
M. Pernici, K. Pilchi and P. van Nieuwenhuizen,
``Gauged $\CN=8$, $d=5$ Supergravity,''
\np{259}{1985}{460}.

\bibitem{rtrunc}
T. T. Tsikas,
``Consistent Truncations of Chiral $\CN=2$, $D=10$ Supergravity
on the Round Five Sphere,''
\cqg{3}{1986}{733}.



\bibitem{rafv}
G. Arutyunov and S. Frolov,
``Some Cubic Couplings in Type IIB Supergravity on $AdS_5\times S^5$ and
Three-point Functions in {\SYMF} at Large $N$,'' hep-th/9907085.


\end{thebibliography}
\end{document}